\documentclass[12pt]{aastex}
\usepackage{emulateapj5}

\newcommand{\um}{$\mu$m}
\newcommand{\kms}{\mbox{\,km\,s$^{-1}$}}

\newcommand{\Msun}{\,$M_{\odot}$}
\newcommand{\Lsun}{\,$L_{\odot}$}
\newcommand{\HII}{\mbox{$\mathrm{H\,{\scriptstyle {II}}}$}}

\newcommand{\MSX}{{\it MSX}}

\newcommand{\Spitzer}{{\it Spitzer}}

\newcommand{\irdcfullname}{MSXDC~G034.43+00.24}
\newcommand{\irdc}{G34.4+0.2}
\newcommand{\cs}{\mbox{CS~(3--2)}}
\newcommand{\hcn}{\mbox{HCN~(4--3)}}
\newcommand{\ceo}{\mbox{C$^{18}$O~(3--2)}}
\newcommand{\sio}{\mbox{SiO~(2--1)}}
\newcommand{\hh}{H$_{2}$ $S$(9) line at 4.694\,\um}

\shorttitle{Massive Protostars \irdcfullname}
\shortauthors{Rathborne et al.}

\begin{document}

\title{Massive Protostars in the Infrared Dark Cloud \irdcfullname}

\author{J. M. Rathborne, J. M. Jackson and E. T. Chambers}
\affil{Institute for Astrophysical Research, Boston University, Boston, MA
02215; rathborn@bu.edu, jackson@bu.edu, etc1@bu.edu}
\author{R. Simon}
\affil{I.Physikalisches Institut, Universit\"at zu K\"oln, 50937 K\"oln, Germany; simonr@ph1.uni-koeln.de}
\and
\author{R. Shipman and W. Frieswijk}
\affil{Kapteyn Astronomical Institute, University of Groningen, and Netherlands Institute for Space Research, P.O. Box 800, 9700 AV Groningen, Netherlands; russ@sron.rug.nl, frieswyk@astro.rug.nl}
%%%%%%%%%%%%%%%%%%%%%%%%%%%%%%%%%%%%%%%%%%

\begin{abstract}
We present a multiwavelength study of the infrared
dark cloud \irdcfullname. Dust emission, traced by millimeter/submillimeter images 
obtained with the IRAM, JCMT, and CSO telescopes, reveals three
compact cores within this infrared dark cloud with masses of 170--800\,\Msun\, 
and sizes $<$ 0.5 pc. \Spitzer\, \mbox{3.6--8.0\,\um}\, images show slightly extended
emission toward these cores, with a spectral enhancement at 4.5\,\um\, that 
probably arises from shocked H$_{2}$. In addition, the broad line widths 
($\Delta V\sim10$\kms) of \hcn\, and \cs, and the detection of \sio, observed with the JCMT and IRAM telescopes, 
also indicate active star formation. \Spitzer\, 24\,\um\, images reveal 
that each of these cores contains a bright, unresolved continuum source; 
these sources are most likely embedded protostars. Their millimeter to mid-IR continuum spectral 
energy distributions reveal very high luminosities, 9000--32,000\,\Lsun.  Because 
such large luminosities cannot arise from low-mass protostars, \irdcfullname\ 
is actively forming massive ($\sim$10\,\Msun) stars.
\end{abstract}

%%%%%%%%%%%%%%%%%%%%%%%%%%%%%%%%%%%%%%%%%%


%%%%%%%%%%%%%%%%%%%%%%%%%%%%%%%%%%%%%%%%%%

\section{Introduction}

Infrared dark clouds (IRDCs) are a distinct class of interstellar gas
cloud discovered in the {\it Infrared Space Observatory} and {\it Midcourse 
Space Experiment } (\MSX)
Galactic plane mid-IR surveys (\citealp{Perault96,Carey98,Hennebelle01}).
IRDCs are identified as dark extinction features seen 
in silhouette against the bright Galactic background at
mid-IR wavelengths.  Preliminary studies show that they
have high densities ($>10^{5}$\,cm$^{-3}$), high column densities 
($\sim10^{23}-10^{25}$ cm$^{-2}$), and low temperatures 
($<25$ K; \citealp{Egan98,Carey98,Carey00}).

It is still unclear what role IRDCs play in star
formation. \cite{Egan98} suggested that IRDCs are
quiescent, isolated molecular clumps, not yet actively
forming stars. Indeed, if a massive star formed inside an IRDC, it would quickly
heat the surrounding natal dust, and thus the IRDC would not remain cold
and dark. The detection of compact millimeter and submillimeter
(\citealp{Lis94,Carey00,Redman03}) continuum sources
toward some IRDCs, however, suggests that they may harbor prestellar cores.
If so, then IRDCs may well play a key role in the very earliest stages of star 
formation. It is important to determine whether these compact cores are sites of
current or future star formation. 

In this {\it Letter}, we use two standard methods to identify sites of active star
formation. The first relies on the interaction between a young stellar 
object (YSO) and its parental molecular cloud. Because YSOs undergo bipolar 
outflows, they interact strongly with the surrounding gas and thus 
can be found by detecting large molecular line widths or shocked gas. One can 
search for shocks by detecting spectral lines that require shocks for excitation 
(e.g., H$_{2}$) or chemical species that require shocks for their formation 
(e.g., SiO).  
The second method relies on the direct detection of the protostar.  Because
protostars are deeply embedded in dusty clouds, they emit most strongly
in the millimeter to IR. Active star-forming cores are thus
associated with luminous, highly reddened, compact sources.

We have used both of these methods to discover 
three distinct sites of active star formation in the IRDC \irdcfullname\, 
(hereafter \irdc). We identified
this IRDC as an extinction feature in the \MSX\, 8\,\um\, Galactic
plane survey \citep{Simon-catalog}. Based on its $^{13}$CO emission from the 
BU-FCRAO Galactic Ring Survey \citep{Simon-grs}, we find that \irdc\, is located
at a kinematic distance of 3.7 kpc (because it is an IRDC, we assume it lies at
the near kinematic distance). 

Superposed against this molecular cloud is the compact \HII\, region
IRAS 18507+0121 \citep{Bronfman96,Molinari96}. \cite{Shepherd04} 
noted a millimeter continuum source (our core MM1) $\sim$40\arcsec\, to the north of this \HII\,
region. From its weak 6 cm emission, they suggest that this millimeter
core contains a deeply embedded B2 protostar.  
Although \cite{Shepherd04} did not note that this millimeter core is in fact part 
of an IRDC, we find that it is indeed embedded within \irdc. In addition,
\cite{Garay04} detected 1.2 mm continuum emission associated with
this cloud in their survey of massive star-forming regions. They note
that one core (our MM3) appears to be a cold ($<$17 K), massive, dense, and quiescent core.

We find strong evidence to support the idea that these two cores, and one other 
within \irdc, are indeed sites of active star formation and
will eventually become massive stars ($M\sim10$ M$_\odot$).
The discovery of active massive star formation in this IRDC suggests that IRDCs may be the earliest
stage in the formation of high-mass stars.

%%%%%%%%%%%%%%%%%%%%%%%%%%%%%%%%%%%%%%%%%%

\section{Observations}

We present both continuum and spectral line data at many wavelengths (see
Table~\ref{obs-summary} for  details). The continuum data cover 
the wavelength range from 1.2 mm to 24\,\um. The millimeter/submillimeter continuum data were 
taken at the Institute de Radioastronomie Millimetrique (IRAM), James Clerk Maxwell Telescope (JCMT), 
and Caltech Submillimeter Observatory (CSO) using either an ``on-the-fly'' or ``scan-mapping'' mode. Standard 
reduction methods within the software packages MOPSIC, CRUSH, and SURF were
used to correct for atmospheric opacity and to remove
atmospheric fluctuations. The millimeter/submillimeter data were flux-calibrated 
using either G34.3 or Uranus.  The IR continuum data were obtained using the
\Spitzer\ Space Telescope.  
The 3.5--8.0\,\um\, images were obtained as part of the Galactic Legacy Infrared Mid-Plane Survey Extraordinaire
(GLIMPSE; \citealp{Benjamin03}) using the Infrared Array Camera (IRAC).
The 24\,\um\, images were obtained using the photometry raster mapping mode of the Multiband Imaging Photometer for \Spitzer\,
(MIPS).
The IRAC data were processed by the GLIMPSE team, and the MIPS data 
were processed by the \Spitzer\, Science Center data processing pipelines. 

The spectroscopic data consist of millimeter/submillimeter molecular line observations obtained with
IRAM and JCMT. Molecular line spectra toward two of the millimeter cores, MM1 and MM3, within \irdc\, were
obtained in \sio, \cs, \hcn, and \ceo. Chopper wheel calibration, pointing, and focus checks
were performed regularly. The data
were reduced using standard methods in the software packages
GILDAS and SPECX.

%%%%%%%%%%%%%%%%%%%%%%%%%%%%%%%%%%%%%%%%%%

\section{Results and Discussion}

\subsection{Identifying Candidate Protostellar Cores}

\begin{figure} 
\epsscale{0.5}
\plotone{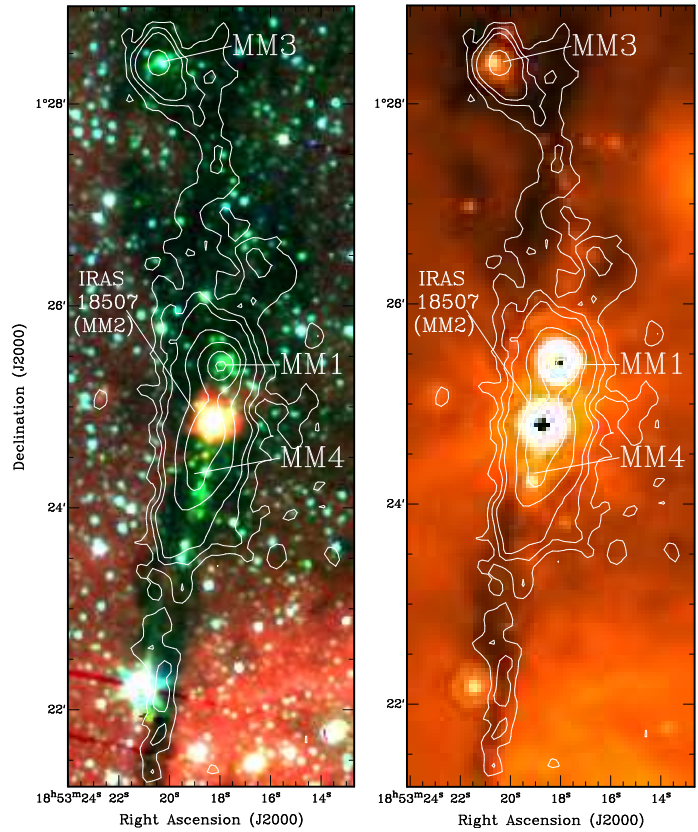}
\caption{\label{irdc43-images}\irdc. {\it Left}: \Spitzer /IRAC three-color image (3.6\,\um\, 
in blue, 4.5\,\um\, in green and 8.0\,\um\, in red) overlaid with IRAM/MAMBO-II 1.2 mm continuum 
emission (contour levels 60, 90, 120, 240, 480, 1200, and 2200 mJy beam$^{-1}$). 
{\it Right}: \Spitzer /MIPS 24\,\um\, image with contours 
of the  IRAM/MAMBO-II 1.2 mm continuum emission (logarithmic color scale, 30 MJy sr$^{-1}$ 
[black] to 200 MJy sr$^{-1}$ [white]). Labelled on this figure are the four millimeter cores.}
\end{figure}

To locate potential sites of star formation, we have imaged \irdc\, in the
millimeter/submillimeter continuum.  Because these wavelengths probe cold
dust emission, millimeter/submillimeter continuum emission can locate dense compact cores, potential
sites of star formation. Our millimeter/submillimeter images of \irdc\, reveal extended, diffuse continuum 
dust emission that matches the morphology of the mid-IR extinction 
extremely well. In addition to this extended emission, we also find four compact cores.
Figure~\ref{irdc43-images} shows the 1.2 millimeter continuum emission 
overlaid on a \Spitzer\, three-color image (3.6\,\um\, in blue, 4.5\,\um\, 
in green, and 8.0\,\um\, in red). One of these cores (MM2) is clearly associated with
the \HII\, region IRAS 18507+0121. The remaining three dust cores (MM1, MM3, and MM4)
are unresolved ($<0.5$ pc) and have 
1.2 mm peak fluxes that range from 0.3 to 2.5 Jy beam$^{-1}$. Table~\ref{source-summary} lists
their coordinates and peak fluxes.

Because the 1.2 mm continuum emission is optically thin, we can reliably estimate 
masses via the expression $ M = F_{\nu} D^{2} /  [\kappa_{\nu} B_{\nu} (T)], $
where F$_{\nu}$ is the observed flux density, {\it D} is the distance,
$\kappa_{\nu}$ is the dust opacity per gram of dust, and B$_{\nu}$(T) is the Planck function at the 
dust temperature. We adopt for $\kappa_{1.2 mm}$ a value of 1.0 cm$^2$ g$^{-1}$ \citep{Ossenkopf94}.
%We estimate $\kappa_{\nu}$ at 1.2mm using 
%$\kappa_{\nu}$ = $\kappa_{230 GHz}$ ($\nu$/230GHz)$^{\beta}$ 
%with $\beta$=2, $\kappa_{230 GHz}$ = 0.01 cm$^{2}$g$^{-1}$ \citep{Ossenkopf94}.
Integrating the 1.2~mm continuum emission above a 3~$\sigma$ level, 
and assuming a dust temperature of $T_{D} = 30$ K and a gas-to-dust
mass ratio of 100, we find that the total mass of the IRDC 
is $\sim$7500\,\Msun.  Gaussian fits to the three cores reveal masses in the 
range 170--800\,\Msun\, (see Table~\ref{source-summary}).
The core masses and sizes are 
similar to massive protostars seen toward other star-forming regions \citep{Brand01}. 

\subsection{Evidence of Outflows and Shocked Gas}

Evidence of kinematic interactions between embedded YSOs and
the molecular gas in these cores can be found in the \Spitzer /IRAC
data. The IRAC three-color 3.5, 4.6, and 8.0\,\um\, image of \irdc\, reveals slightly
extended emission toward all three millimeter cores, with a 
relative enhancement at 4.5\,\um\, (displayed as green). 
Although highly extincted stars can appear green in these three color images due to 
the flat extinction curve between 4.5 and 8.0\,\um\, \citep{Remy05}, bright, extended 4.5\,\um\,
emission has been found to trace shocked gas in star-forming 
regions \citep{Marston04,Noriega04}. This enhanced 4.5\,\um\, emission 
is thought to arise from shock-excited spectral lines, probably the \hh.
Because these cores show evidence of shocked gas, they are probably forming
stars.

\begin{figure}
\epsscale{0.5}
\plotone{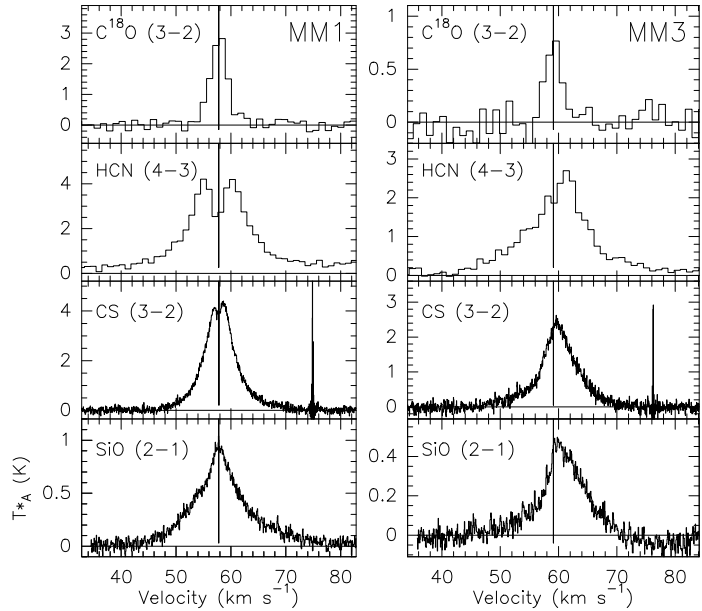}
\caption{\label{lines}IRAM and JCMT molecular line emission toward two of the millimeter cores. 
The solid vertical line marks the central velocity of the core [as traced by
the optically thin \ceo\, emission]. Note the broad line widths 
in the \hcn\, and \cs\, spectra.}
\end{figure}

Further evidence comes from our IRAM and JCMT molecular 
line spectra toward two of these cores (Fig.~\ref{lines}).
The high-density tracing molecular lines, \cs\ and \hcn, show broad 
line widths ($\Delta V\sim$10\,\kms). Such broad line widths are often interpreted 
as tracing protostellar outflows \citep{Moriarty95}. Because SiO is thought to form as 
the result of the shock destruction of dust grains, the \sio\, line indicates strong 
shocks and is therefore a reliable indicator of star formation 
\citep{Bachiller91,Martin92,Zhang95,Gibb04}. The strong SiO line, combined with 
broad lines of the molecular high-density tracers, provides further evidence 
that active star formation is occurring within these cores.

\subsection{Embedded Protostars}

A more direct indicator of active star formation is the detection of embedded
protostars via their millimeter to IR continuum emission.  We have detected embedded
protostars toward the three cores using 
new \Spitzer\, 24\,\um\, data. The 24 \um\, image of \irdc\, shows that, 
while the IRDC itself remains dark, 
the active cores each contain unresolved ($<$6\arcsec), bright
24\,\um\, continuum sources (Fig.~\ref{irdc43-images}). 

\begin{figure}
\epsscale{0.5}
\plotone{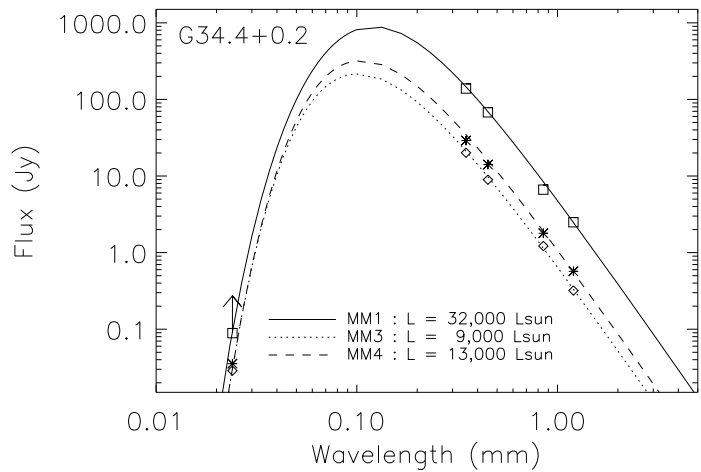}
\caption{\label{sed}Broadband continuum SED for the three cores. Included on
this plot are peak fluxes at 24\,\um, 350\,\um, 450\,\um, 850\,\um, and 1.2~mm. 
The curves are graybody fits to the data, which yield 
values of $\beta$$\sim$1.8, $\tau_{250}$$\sim$0.25, and
$T_{D}$$\sim$33\,K (15\arcsec\, source diameter; see also Table~\ref{source-summary}). 
The derived bolometric luminosities are labeled for each core. MM1 saturates the MIPS array;
hence, the quoted 24\,\um\, flux is a lower limit.}
\end{figure}

We can determine the luminosity of the embedded protostars from their millimeter to IR 
broadband spectral energy distributions (SEDs). Using the peak fluxes from 
all of our continuum data in the mid-IR (24 \um), submillimeter (350, 450, and 850 \um),
and millimeter (1.2 mm), we can construct broadband SEDs and fit them to standard
graybody curves (Fig.~\ref{sed}). For simplicity, we assume that the cores 
are isothermal. In our fits, the emissivity index ($\beta$), 
optical depth at 250\,\um\, ($\tau_{250}$), and dust temperature ($T_{D}$) are 
free parameters. 
For a source size of 15\arcsec, these fits yield 
reasonable values for these parameters: $\beta$$\sim$1.8, $\tau_{250}$$\sim$0.25, and
$T_{D}$$\sim$33\,K (see Table~\ref{source-summary}). The SEDs also provide good estimates for
the bolometric luminosity, which is simply the integral under the SED fit. Because
these sources are slightly resolved, fits to the SEDs using the peak fluxes will yield lower
limits to the actual luminosity. 
We find that the three protostellar cores in \irdc\, have bolometric luminosities
($L_{bol}$) of \mbox{9000--32,000\,\Lsun}.

\subsection{Large Luminosities: High-Mass Protostars}

Theoretical studies of protostellar evolution suggest that low-mass stars 
(M$<$2\,\Msun) never achieve luminosities $>$~100\,\Lsun\, in their pre--main-sequence
evolution \citep{Iben65}. High-mass 
protostars (M$>$8\,\Msun), however, evolve at essentially constant luminosity on 
their path from protostar to main-sequence star \citep{Iben65,Palla90}. Thus, 
the protostellar 
luminosity is a good indicator of the eventual main-sequence luminosity and, hence, 
stellar mass. Protostellar luminosities of $\sim$10,000\,\Lsun, therefore, 
correspond to early B stars with masses of $\sim$10\,\Msun\, \citep{Panagia73}. Because 
\irdc\, contains three sites of active star formation with $L_{bol}$ 
$\sim$10,000\,\Lsun\, or more, we conclude that each of these sites contains a massive
protostar. Specifically, we find corresponding spectral types of O9.5 for MM1 and B0.5 for
both MM3 and MM4 \citep{Panagia73}.

%Empirical studies of massive protostars support our conclusion. \cite{Sridharan02}
% have shown a good correlation between bolometric luminsoity and the
%molecular gas mass associated with a massive protostar.  For $L_{bol}\sim$10,000\,\Lsun,
%their relation suggests a total molecular core mass of M$\sim>$100\,\Msun.
%For reasonable star formation efficiencies (30\,\%, \citealp{Lada-review}), such large molecular 
%core masses produce massive stars M$>$10\,\Msun.  Consequently,
%the protostars in \irdc\, will almost certainly evolve into high mass stars.
 
\subsection{The Role of IRDCs in Star Formation}

The discovery of active massive star formation in \irdc\, shows that the earliest
stages of massive star formation may well take place in IRDCs. 
Because low-mass stars form within Bok globules, high-mass stars probably form
within a higher mass equivalent. We speculate
that IRDC cores are indeed these high-mass counterparts to Bok globules.
If IRDCs host the earliest stages of high-mass star formation, 
and then at some point they will spawn OB stars and grow warm.
Indeed, the masses and sizes of IRDCs ($M$$\sim$1000\,\Msun, sizes approximately a few parsecs)
are similar to those of warm molecular clumps associated with massive star-forming
regions \citep{Garay99,Lada-review}. Moreover, such warm clumps often produce
multiple massive stars. The fact that \irdc\, is forming three massive stars and
has already formed a B0.5 star associated with the \HII\, region IRAS 18507+0121 
\citep{Shepherd04} suggests that IRDCs can also form several massive stars almost
simultaneously. To test this hypothesis, we are currently searching for more 
examples of active star formation within a larger sample of IRDCs. 

%%%%%%%%%%%%%%%%%%%%%%%%%%%%%%%%%%%%%%%%%%
\section{Conclusions}

In order to search for regions of active star formation in IRDCs, 
we have performed a multiwavelength study of the IRDC \irdc.
We identified three potential star-forming cores by finding compact millimeter/submillimeter 
continuum sources. These cores show
enhanced 4.5\,\um\, emission, which is thought to trace shocked gas in star-forming regions.  
Broad molecular line widths of CS and HCN provide additional evidence of
star formation toward two cores. Each of these cores also has strong, broad
lines of SiO, a species that preferentially forms in shocks.

We have also directly detected bright, highly reddened, compact sources, 
presumably protostars, in all three cores. The broadband SEDs
of these cores, from millimeter to mid-IR wavelengths, combined with the
known kinematic distance, provide a good estimate of their bolometric 
luminosities, 9000--32,000 L$_\odot$.  Both theoretical and empirical studies
suggest that such large luminosities can only arise from high-mass
protostars. Hence, \irdc\, contains three active sites of 
star formation, which will eventually produce massive stars
with $M\sim10$\,M$_\odot$.  This study suggests that IRDCs play
an important role in the early stages of massive star formation.
 
%%%%%%%%%%%%%%%%%%%%%%%%%%%%%%%%%%%%%%%%%%

\acknowledgments
The authors gratefully acknowledge funding support through NASA grant NNG04GGC92G. 
This work is based in part on observations made with the {\it Spitzer Space Telescope}, 
which is operated by the Jet Propulsion Laboratory, California Institute of Technology 
under NASA contract 1407. Support for this work was provided by NASA through 
contract 1267945 issued by JPL/Caltech. The JCMT is operated by JAC, Hilo, 
on behalf of the parent organizations of the Particle Physics and Astronomy Research 
Council in the UK, the National Research Council in Canada, and the Scientific Research 
Organization of the Netherlands. IRAM is supported by INSU/CNRS (France), MPG (Germany), 
and IGN (Spain). The CSO telescope is operated by Caltech under a contract from the 
National Science Foundation (NSF). We would like to thank the GLIMPSE team (PI: E. 
Churchwell) for providing the IRAC images.

%%%%%%%%%%%%%%%%%%%%%%%%%%%%%%%%%%%%%%%%%%

%%%%%%%%%%%%%%%%%%%%%%%%%%%%%%%%%%%%%%%%%

\begin{table}
\centering
\caption{\label{obs-summary}Summary of observations}
\begin{tabular}{ccccccc}
\tableline \tableline
Telescope &Instrument&Wavelength/ &Date & Angular&1$\sigma$ rms \\
          &         & Spectral line &  & Resolution (\arcsec) & noise   \\
\tableline
IRAM 30\,m & MAMBO-II  &  1.2 mm         & 2004 Feb      & 11    & 10 mJy\\
JCMT 15 m  & SCUBA     & 850, 450\,\um   & 2004 Sept     & 15, 8 & 100, 300 mJy\\
CSO 10\,m  & SHARC-2   & 350\,\um        & 2005 Apr      & 8     & 500 mJy\\
\Spitzer   & MIPS      & 24\,\um         & 2004 Oct      & 6     & 70 $\mu$Jy\\
           & IRAC      & 3.6, 4.5, 8.0\,\um   & 2003--2004 & 2  & 23, 27, 78 $\mu$Jy\\
\tableline
IRAM 30\,m & C150 Rx   &\cs             & 2004 Nov       & 26\tablenotemark{a}    & 0.13 K \\
           & A100 Rx   &\sio            & 2004 Nov       & 28  & 0.06 K \\
JCMT 15\,m & B3 Rx     &\hcn, \ceo      & 2004 Apr       & 14  & 0.07, 0.13 K\\
\tableline
\end{tabular}\vspace{-0.5cm}
\tablenotetext{a}{Mapped and then smoothed to this resolution} 
\end{table}

\begin{table}
\centering
\caption{\label{source-summary}Parameters for the 1.2 mm cores within \irdc.}
\begin{tabular}{cccccccccccccc}
\tableline \tableline
Source\tablenotemark{a} & \multicolumn{2}{c}{Coordinates}  &&\multicolumn{5}{c}{Peak Flux} &  & \multicolumn{4}{c}{SED fit parameters} \\
\cline{2-3} \cline{5-9} \cline{11-14}
    & $\alpha$ & $\delta$   && 1.2mm & 850\um & 450\um & 350\um & 24\um & M$_{1.2mm}$\tablenotemark{b}  & $\beta$ & $\tau_{250}$ &T$_{D}$ &  $L_{bol}$ \\
    & (J2000)  & (J2000)    && (Jy)   &  (Jy)    & (Jy)       & (Jy)       & (Jy)      & (\Msun) &  & &(K) &  (\Lsun) \\
\tableline
MM1 & 18 53 18.0 & 01 25 24  &&  2.5   & 6.6     & 70      & 140    & $>$0.09      & 800 & 1.8 & 0.53 & 34 &   32,000\\
MM3 & 18 53 20.4 & 01 28 23  &&  0.3   & 1.2     & 10      & 20    & 0.03    & 200 & 1.8 & 0.08 & 32 &  ~9,000\\
MM4 & 18 53 19.0 & 01 24 08  &&  0.6   & 1.8     & 15      & 30     & 0.04     & 170 & 1.8 & 0.13 & 32 &  12,000\\
\tableline
\end{tabular}\vspace{-0.5cm}
\tablenotetext{a}{The cores are named in order of peak 1.2 mm flux (determined from gaussian fits). IRAS 18507+0121 corresponds to our core MM2.}
\tablenotetext{b}{Calculated from a gaussian fit to the 1.2 mm emission associated with the core (excluding the
underlying emission from the IRDC).}
\end{table}

\end{document}